\documentclass[twocolumn]{aastex631}

\usepackage{graphicx}	
\usepackage{multirow}

\usepackage{amsmath}

\usepackage{url}
\usepackage{hyperref}

\shorttitle{Galaxy quenching across the Cosmic Web}
\shortauthors{Nandi, and Pandey}

\def\refjnl#1{{\rm#1}}
\def\apjl{\refjnl{ApJ Letters}}                
\def\apjs{\refjnl{ApJS}}               
\def\ao{\refjnl{Appl.~Opt.}}           
\def\aap{\refjnl{A\&A}}                

\usepackage{hyperref}

\begin{document}

\title{Galaxy quenching across the Cosmic Web: disentangling mass and environment with SDSS DR18}

\correspondingauthor{Biswajit Pandey}
\email{biswap@visva-bharati.ac.in}

\author{Anindita Nandi}
\affiliation{Department of Physics, Visva-Bharati University, Santiniketan, 731235, India}

\author{Biswajit Pandey}
\affiliation{Department of Physics, Visva-Bharati University, Santiniketan, 731235, India}

\begin{abstract}
We investigate the influence of large-scale cosmic web environments on galaxy quenching using a volume-limited, stellar mass-matched galaxy sample from SDSS DR18. Galaxies are classified as residing in sheets, filaments, or clusters based on the eigenvalues of the tidal tensor derived from the smoothed density field. The quenched fraction increases with stellar mass and is highest in clusters, intermediate in filaments, and lowest in sheets, reflecting the increasing efficiency of environmental quenching with density. A flattening of the quenched fraction beyond $\log_{10}(M_\star/M_\odot) \sim 10.6$ across all environments signals a transition from environment-driven to mass-driven quenching. In contrast, the bulge fraction continues to rise beyond this threshold, indicating a decoupling between star formation suppression and morphological transformation. At the high-mass end ($\log_{10}(M_\star/M_\odot) \gtrsim 11.5$), both quenched and bulge fractions bifurcate, increasing in clusters but declining in sheets, suggesting a divergent evolutionary pathway where massive galaxies in sheets retain cold gas and disk-like morphologies, potentially sustaining or rejuvenating star formation. The AGN fraction also increases with stellar mass and is somewhat higher in sheets than in clusters, indicating enhanced AGN activity in low-density, gas-rich environments. The high-mass trends are independently corroborated by our analysis of specific star formation rate, $(u-r)$ colour, concentration index, and D4000 in the stellar mass-density plane, which show that massive galaxies in sheets remain bluer, younger, more star-forming, and structurally less evolved than their cluster counterparts. Our results highlight the cosmic web as an active driver of galaxy evolution.
\end{abstract}



\section{Introduction}
\label{sec:intro}

The large-scale structure of the Universe, the cosmic web \citep{bond96} forms the scaffolding upon which galaxies assemble, evolve, and ultimately quench. This intricate network of voids, sheets, filaments, and clusters arises from the anisotropic gravitational collapse of primordial density perturbations, and has been observed across decades of galaxy surveys, including CfA \citep{geller89}, LCRS \citep{schectman96}, SDSS \citep{york00}, 2dFGRS \citep{colless01}, GAMA \citep{driver11}, 2MRS \citep{huchra12}, KiDS \citep{kuijken19}, and DESI \citep{desi25}. Both observational studies \citep{gregory78, einasto80, bharad00, pandey05, tempel14} and cosmological simulations \citep{springel05, calvo10, schaye15, vogelsberger14, cautun14, libeskind18} have consistently revealed that galaxies are not randomly distributed in space but instead trace the geometry of this web-like structure. 

Recent advances have deepened our understanding of how the cosmic web environment influences galaxy evolution. Filaments, the most extended and coherent components of the web \citep{bharad04, pandey11}, not only serve as channels for anisotropic matter flows but also host a significant fraction ($\sim 40\%-50\%$) of the Universe’s baryons in the form of the Warm-Hot Intergalactic Medium (WHIM) \citep{cen99, dave01, tuominen21, galarraga21}. Observational detections of WHIM through X-ray absorption and Sunyaev–Zeldovich effects \citep{nicastro18, singari20} support its role as a diffuse gas reservoir regulating galaxy growth via cold gas accretion. Galaxies embedded in sheets and filaments can draw upon this reservoir, sustaining star formation through directed gas inflows \citep{chen17, pandey20, singh20, das22, das23, hoosain24}. In contrast, clusters formed at the intersections of filaments represent the densest and most hostile environments, characterized by high-temperature intracluster media, frequent interactions, and extreme quenching processes \citep{gunn72, roediger07, ruggiero17}.

The spatial and dynamical hierarchy of the cosmic web also regulates the assembly histories of dark matter halos and their baryonic contents. Matter flows from voids to walls, then into filaments, and finally accumulates in clusters \citep{aragon10, cautun14, ramachandra15, wang24}. This progression shapes halo and galaxy properties beyond mass alone. The phenomenon of halo assembly bias, the dependence of halo clustering on formation history \citep{gao07, croton07, dalal08, salvador24} has an analogous manifestation in the baryonic sector, galaxy assembly bias, wherein galaxies of similar mass but different assembly histories exhibit different observable properties. These assembly histories are intimately linked to the galaxy’s position within the cosmic web, suggesting that environment can imprint lasting signatures on galaxy evolution, even at fixed stellar mass.

A pivotal phase in this evolutionary trajectory is galaxy quenching, the cessation of star formation which transforms blue, star-forming galaxies into red, quiescent systems. The bimodality observed in galaxy colour, morphology, and star formation rate distributions \citep{strateva01, blanton03, balogh04a, baldry04,pandey20a} is widely interpreted as the outcome of this transition. Observations indicate a sharp decline in the global star formation rate density from $z\sim1$ to the present epoch \citep{madau96, madau14}, accompanied by the growth of the red sequence \citep{bell04, faber07}. This transformation is thought to arise from a combination of secular and environmental mechanisms, whose relative importance likely varies across the cosmic web.

Theoretical models and hydrodynamical simulations suggest that galaxies in low-mass halos ($M_{\mathrm halo} \lesssim 10^{12} \, M_\odot$) can efficiently accrete cold gas via filamentary inflows, sustaining star formation \citep{birnboim03, dekel06, keres05, gabor10}. However, in more massive halos, a virial shock develops, heating the gas to high temperatures and inhibiting cold mode accretion, a process known as mass quenching \citep{binney04, birnboim07}. To maintain quenching, additional heating mechanisms such as AGN feedback \citep{croton06, bower06, somerville08} are required to prevent the hot halo gas from cooling and re-initiating star formation. Other secular processes such as morphological quenching \citep{martig09}, bar-driven gas depletion \citep{masters10}, and angular momentum suppression \citep{peng20} can also inhibit star formation in the absence of external triggers.

Environment-driven mechanisms, on the other hand, depend on the surrounding galaxy density and dynamics. These include ram-pressure stripping \citep{gunn72}, starvation \citep{larson80, kawata08}, strangulation \citep{balogh00}, harassment \citep{moore96}, and tidal interactions or mergers that dynamically heat or deplete cold gas reservoirs \citep{toomre72, mihos96, hopkins13, renaud15, moreno21}. Such processes are most effective in high-density environments like clusters but can also operate in filaments and groups, where galaxies may be pre-processed before accreting onto clusters \citep{fujita04, darvish17a, sarron19}. Thus, the cosmic web acts as both a conduit and a regulator facilitating cold gas inflows in some environments while suppressing them in others.

Despite significant progress, a comprehensive understanding of how quenching processes unfold across the different structures of the cosmic web remains elusive. Observational studies leveraging SDSS data have found gradients in quenched fractions, bulge dominance, and AGN activity across sheets, filaments, and clusters \citep{argudo16, darvish17b, kuutma17, pandey20, malavasi22}. High-resolution simulations such as IllustrisTNG \citep{nelson19} and EAGLE \citep{schaye15} support these trends, revealing environment-dependent variations in gas accretion rates, star formation efficiencies, and galaxy morphology \citep{voort17, tacchella19, nelson19, kraljic20,  wright21, pandey24}. However, many of these studies either do not disentangle the roles of mass and environment, or rely on local density as a proxy for environment, overlooking the topological distinctions encoded in the cosmic web.

In this study, we aim to address these gaps by analyzing a volume-limited sample of galaxies from the Sloan Digital Sky Survey (SDSS) Data Release 18. We classify galaxies into distinct cosmic web environments: voids, sheets, filaments, and clusters using a Hessian-based approach that utilizes the eigenvalues of the deformation tensor to characterize the local topology. After removing the sparsely populated void galaxies, we construct stellar mass-matched subsamples across the three environments, each containing same number of galaxies. 
We focus our investigation on three key diagnostics of galaxy quenching: the quenched fraction, the bulge fraction, and the AGN fraction. By examining how these quantities vary as a function of stellar mass across the sheet, filament, and cluster environments, we aim to disentangle the respective roles of internal (mass-driven) and external (environment-driven) mechanisms in suppressing star formation. 

Through this analysis, we aim to advance our understanding of the complex interplay between stellar mass, local density, large-scale environment, and the cessation of star formation. Specifically, we investigate how the quenched fraction of galaxies varies with stellar mass across sheets, filaments, and clusters, examine the influence of bulge formation and AGN activity in driving quenching within these environments, and assess how local density modulates the relative importance of mass-driven and environment-driven quenching processes. To gain a more comprehensive picture of galaxy transformation across environments, we further analyze specific star formation rate, $(u-r)$ colour, concentration index, and D4000 strength in the stellar mass-density plane. These complementary diagnostics provide independent insights into the star formation activity, morphology, and stellar population age of galaxies, allowing us to trace the physical pathways of quenching and morphological evolution beyond simple thresholds in mass or environment. By linking these internal properties to the structural context of the cosmic web, we explore whether patterns in star formation suppression reflect deeper variations in gas accretion efficiency, feedback strength, or dynamical history across different cosmic environments. In doing so, we seek to clarify the role of the cosmic web not merely as a backdrop for galaxy evolution, but as an active agent shaping its trajectory through a combination of dynamical, hydrodynamical, and feedback processes.

Throughout this work, we adopt a flat $\Lambda$CDM cosmology with parameters $\Omega_{m0} = 0.315$, $\Omega_{\Lambda0} = 0.685$, and $h = 0.674$ \citep{planck20}.

The structure of the paper is as follows: Section~\ref{sec:data_method} describes the data and methodology, Section~\ref{sec:results} presents the main results, and Section~\ref{sec:conclusion} summarizes our conclusions.


\section{Data and method of analysis}
\label{sec:data_method}

\subsection{Galaxy sample and derived properties}

We construct our galaxy sample from the Sloan Digital Sky Survey Data Release 18 (SDSS DR18; \citealt{almeida2023}), the first public data release of SDSS-V. The SDSS~\citep{york00} has mapped large volumes of the nearby universe using the dedicated 2.5-meter Sloan Telescope~\citep{gunn2006}, providing precise photometric and spectroscopic data.

We retrieve data through the \texttt{CasJobs} interface, using Structured Query Language (SQL) queries. Redshift and photometry are obtained from the \texttt{SpecObj} and \texttt{PhotoObj} tables. Stellar masses and specific star formation rates (sSFRs) are drawn from the \texttt{StellarMassFSPSGranWideDust} table, where physical properties are derived from spectral energy distribution (SED) fitting using composite stellar population models generated with the Flexible Stellar Population Synthesis (FSPS) code~\citep{conroy2008}. Galaxy morphology is quantified using the concentration index, defined as the ratio $r_{90}/r_{50}$, where $r_{90}$ and $r_{50}$ are the Petrosian radii enclosing 90\% and 50\% of the $r$-band flux, respectively~\citep{shimasaku2001}. Galaxies with $r_{90}/r_{50} > 2.6$ are classified as bulge-dominated \citep{strateva01}. We obtain measurements of the $4000\,\AA$ break strength (D4000) from the \texttt{galSpecIndx} table. This spectral feature serves as a reliable tracer of the luminosity-weighted age of a galaxy's stellar population, with larger values typically indicating older, more evolved systems~\citep{balogh99}. AGN are identified using the \texttt{galSpecExtra} table, which is derived from the MPA-JHU spectroscopic galaxy catalogue ~\citep{kauffmann03, kauffmann03a, brinchmann04} and classifies galaxies based on the BPT diagram~\citep{baldwin81}. In this classification scheme, AGN are flagged with a value of 4, corresponding to narrow-line AGN with high signal-to-noise ratios. This conservative selection ensures a clean AGN sample by excluding composite galaxies (flag 3) and systems dominated by low-ionization nuclear emission-line regions (LINERs). Applying this criterion to our volume-limited galaxy sample yields a total of 4,137 AGN.

\begin{figure*}
    \centering
    \includegraphics[width=0.65\textwidth]{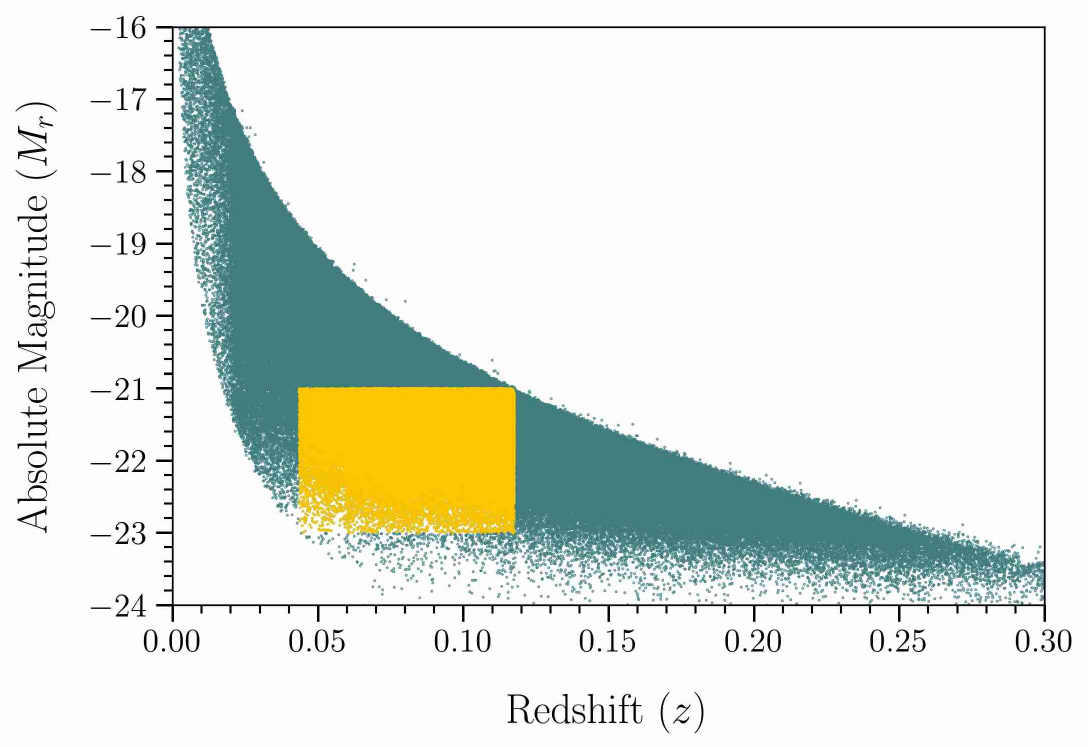}
    \caption{Distribution of galaxies in the redshift (z)-absolute magnitude ($M_r$) plane. The yellow region marks the volume-limited sample used in this work, defined by $-23 \leq M_r \leq -21$ and $0.0434 \leq z \leq 0.1175$.}
    \label{fig:mr_z_plane}
\end{figure*}

We construct a volume-limited sample by selecting galaxies with extinction- and $k$-corrected $r$-band absolute magnitudes in the range $-23 \leq M_r \leq -21$, corresponding to redshifts between $0.0434 \leq z \leq 0.1175$. This selection ensures completeness across environments and stellar masses. The resulting sample is illustrated in \autoref{fig:mr_z_plane}. To remove extreme outliers, we apply quality cuts to several key galaxy properties, listed in Table~\ref{tab:galprops_range}, yielding a final sample of 88,579 galaxies.

Our choice of a volume-limited galaxy sample is driven by the dual requirement of robust cosmic web identification and statistically meaningful comparisons across environments at the high-mass end. While we acknowledge that the chosen absolute-magnitude selection can bias against low-mass quiescent galaxies, our primary scientific focus is on the quenching pathways of massive systems, for which this bias is minimal. Crucially, cosmic web classification relies on the spatial distribution of galaxies to reconstruct a smooth and well-sampled density field. Weighting schemes such as $1/V_{\rm max}$ can correct number statistics but cannot recover the spatial information of galaxies absent from the sample, making them unsuitable for Hessian-based web identification.

We therefore adopt a volume-limited sample that balances survey volume and tracer number density, ensuring a uniform sampling of the underlying density field and preserving the coherence of large-scale structures. We explicitly tested alternative volume-limited selections at lower and higher redshifts. A fainter, low-redshift sample, while better sampling low-mass galaxies, probes too small a volume and yields insufficient statistics for massive galaxies after mass matching across the cosmic web environments. Conversely, a brighter, high-redshift sample significantly lowers the tracer density, requiring excessive smoothing that erases the filamentary and sheet-like features of the cosmic web. Our fiducial sample thus represents an optimal compromise, enabling reliable cosmic web reconstruction while retaining sufficient statistics to robustly characterize quenching, morphology, and AGN activity in massive galaxies across environments.

\begin{deluxetable}{ccc}
\tablecaption{Selection criteria used to define the final galaxy sample. \label{tab:galprops_range}}
\tablewidth{0pt}
\tablehead{
\colhead{Property} & \colhead{Minimum} & \colhead{Maximum}
}
\startdata
$(u-r)$ colour & 0.5 & 4.5 \\
$\log_{10}(M_\star/M_\odot)$ & 10.0 & 12.0 \\
$\log_{10}(sSFR/yr^{-1})$ & $-31$ & $-9.2$ \\
Concentration index ($r_{90}/r_{50}$) & 1.5 & 4.5 \\
D$4000$  & 0.8 & 2.5 \\
\enddata
\end{deluxetable}

\subsection{Quenched galaxy definition}

\begin{figure*}
    \centering
    \includegraphics[width=0.65\textwidth]{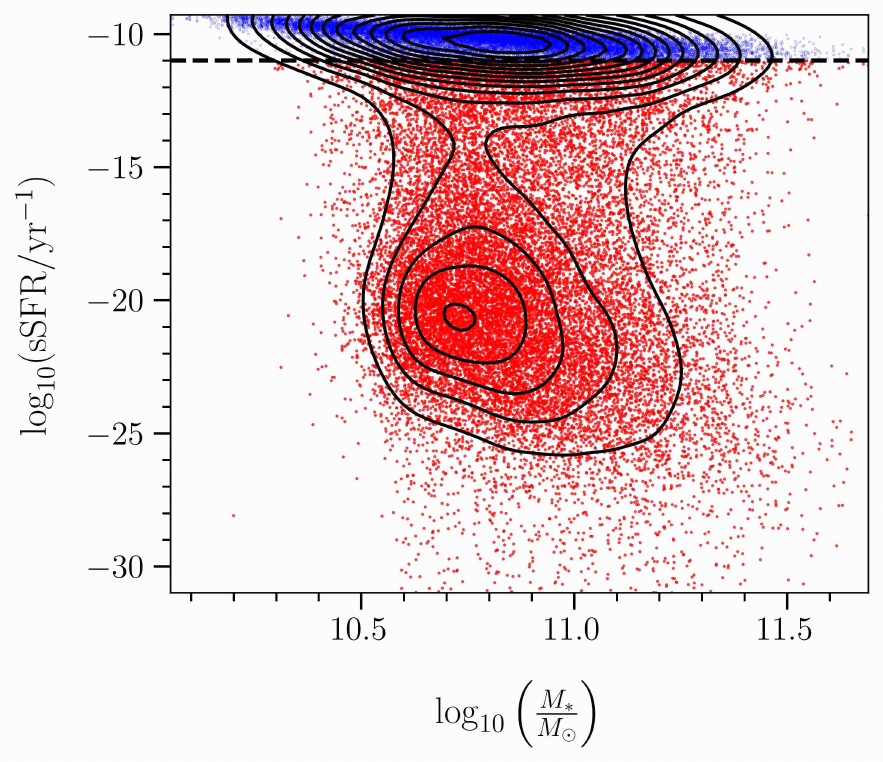}
    \caption{This shows the definition of quenched galaxies in the $\log_{10}(M_\star/M_\odot)$-$\log_{10}(\mathrm{sSFR}/\mathrm{yr}^{-1})$ plane for our volume limited sample. The horizontal dashed line marks the quenching threshold of $\log_{10}(\mathrm{sSFR}/\mathrm{yr}^{-1}) = -11$. Iso-density contours indicate the distribution of galaxies in this plane.}
    \label{fig:ssfr_mass_cut}
\end{figure*}

We define quenched galaxies as those with $\log_{10}(\mathrm{sSFR/yr}^{-1}) \leq -11$, a threshold widely adopted in both observations and simulations \citep{fontanot09, wetzel12, schawinski14, sherman20, donnari21} to separate actively star-forming systems from galaxies with strongly suppressed star formation. \autoref{fig:ssfr_mass_cut} shows the distribution of galaxies in the $\log_{10}(M_\star/M_\odot)$-$\log_{10}(\mathrm{sSFR}/\mathrm{yr}^{-1})$ plane, along with isodensity contours and the quenching threshold. While the sSFR distribution in our sample exhibits a pronounced minimum at lower values (around $\log_{10}(\mathrm{sSFR/yr}^{-1}) \sim -14$ in \autoref{fig:ssfr_mass_cut}), adopting such a stringent cut would isolate only the most extremely passive systems and exclude galaxies undergoing partial or recent quenching. Our science goal is not to distinguish between degrees of quiescence, but to capture the full population of galaxies whose star formation has been significantly suppressed, including transitional systems that are particularly sensitive to environmental regulation. Using a more conservative threshold therefore preserves statistical power, avoids biasing the analysis toward only the most extreme objects, and enables direct comparison with previous studies.

\begin{figure}
    \centering
    \includegraphics[width= \columnwidth]{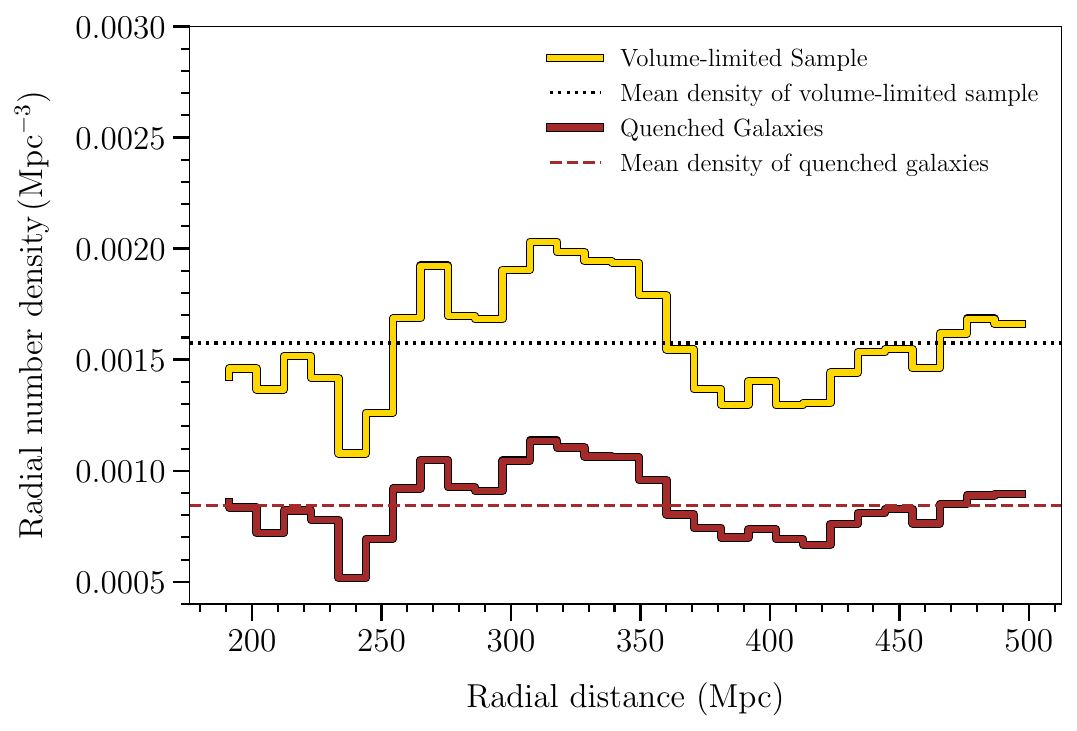}
    \caption{This shows the comoving number density as a function of radial distance for the full volume-limited sample (yellow) and for quenched galaxies (brown). Both curves exhibit qualitatively similar behaviour across the surveyed redshift range. The absence of a systematic decline in the quenched population toward larger distances indicates that quiescent galaxies are not preferentially lost due to selection effects. Residual fluctuations reflect genuine large-scale structure variations rather than incompleteness.}
    \label{fig:radquench}
\end{figure}

To further verify that our quenched population is not affected by redshift-dependent selection biases, we examine the comoving number density of quenched galaxies as a function of radial distance within the survey volume (\autoref{fig:radquench}). For comparison, we also show the corresponding number density of the full volume-limited sample. The two curves display qualitatively similar radial behaviour, with no systematic decline in the quenched population toward larger distances. This demonstrates that quiescent galaxies are not preferentially excluded at higher redshift despite their typically higher mass-to-light ratios. The residual fluctuations observed in both distributions trace genuine large-scale structure variations rather than observational incompleteness. We therefore conclude that our adopted sSFR threshold and volume-limited selection do not introduce a redshift-dependent bias in the quenched fraction.

\subsection{Cosmic web classification}

To classify the large-scale environment of galaxies within the cosmic web, we adopt a Hessian-based framework that identifies morphological structures: voids, sheets, filaments, and clusters based on the eigenvalues of the deformation tensor computed from the smoothed galaxy density field ~\citep{hahn07, forero09}. This technique enables a geometric decomposition of the cosmic web, capturing the underlying gravitational dynamics shaping galaxy environments.

We begin by interpolating the discrete galaxy distribution onto a $256^3$ grid using the Cloud-in-Cell (CIC) scheme to obtain a continuous three-dimensional density field, $\delta(\mathbf{x}) = (\rho(\mathbf{x}) - \bar{\rho})/\bar{\rho}$, where $\bar{\rho}$ is the mean galaxy density. To isolate large-scale structures and suppress small-scale fluctuations, we apply a Gaussian smoothing filter with a characteristic scale of $8\,\mathrm{Mpc}$ which is close to the mean intergalactic separation ($8.57\,\mathrm{Mpc}$) of our volume limited sample.

To identify the cosmic web environment of galaxies, we begin by computing the gravitational potential associated with the smoothed matter density field. This is done in Fourier space using the relation
\begin{equation}
  \hat{\Phi}(\mathbf{k}) = \hat{G}(\mathbf{k}) \, \hat{\rho}(\mathbf{k}),
  \label{eq:phik}
\end{equation}

where $\hat{\Phi}(\mathbf{k})$ is the Fourier transform of the gravitational potential $\Phi(\mathbf{x})$, and $\hat{\rho}(\mathbf{k})$ is the Fourier-transformed, smoothed density contrast field. The function $\hat{G}(\mathbf{k})$ denotes the Green's function of the Laplace operator in Fourier space.

After computing $\hat{\Phi}(\mathbf{k})$, we perform an inverse Fourier transform to recover the gravitational potential $\Phi(\mathbf{x})$ in real space. The deformation tensor, which is the Hessian of the potential field, is then obtained by numerically evaluating the second-order partial derivatives of $\Phi(\mathbf{x})$. The eigenvalues of this tensor at each spatial location provide a dynamical basis for classifying the cosmic web into voids, sheets, filaments, and clusters.

We then construct the deformation tensor, defined as the Hessian of the gravitational potential
\begin{equation}
  T_{\alpha\beta} = \frac{\partial^2 \Phi}{\partial x_\alpha \partial x_\beta}, \quad \alpha, \beta \in \{x, y, z\},
\end{equation}
which encapsulates the local tidal field and encodes the curvature of the gravitational potential at each grid point. The three eigenvalues $\lambda_1 > \lambda_2 > \lambda_3$ of this tensor are computed numerically.

The cosmic web environment at each location is classified based on the number of eigenvalues. Specifically:
\begin{itemize}
    \item Void: $\lambda_1 < 0$, $\lambda_2 < 0$, $\lambda_3 < 0$
    \item Sheet: $\lambda_1 > 0$, $\lambda_2 < 0$, $\lambda_3 < 0$
    \item Filament: $\lambda_1 > 0$, $\lambda_2 > 0$, $\lambda_3 < 0$
    \item Cluster: $\lambda_1 > 0$, $\lambda_2 > 0$, $\lambda_3 > 0$
\end{itemize}

Each galaxy in our sample is then assigned a cosmic web classification corresponding to the grid cell it occupies. This eigenvalue-based classification reflects the anisotropic gravitational influence experienced by galaxies in different environments. 

\subsection{Local density estimation}

To quantify the local environment around each galaxy, we employ a well-established nearest-neighbour technique ~\citep{casertano85}, which estimates local density based on the proximity of neighbouring galaxies. We use the fifth-nearest neighbour density, $\rho_5$, defined as:

\begin{equation}
\rho_k = \frac{k - 1}{V(r_k)},
\end{equation}

\noindent where $r_k$ is the distance to the $k^\mathrm{th}$ nearest neighbour, and $V(r_k)$ denotes the volume of the sphere enclosing those neighbours.

The choice of the nearest-neighbour parameter $k$ in local density estimators is not unique and represents a trade-off between sensitivity to small-scale structure and statistical stability \citep{keller85, garcia08, papani21}. We adopt $k=5$, a commonly used value in observational studies \citep{balogh04b, mateus04, einasto05, peng10, bolzonella10, schaefer17, santucci23}, as a representative scale that probes the local galaxy environment while mitigating shot noise. To assess the robustness of our results, we have repeated our analysis using $k=3$ and $k=7$. Very small values of $k$ (e.g. $k=2$) yield noisy density fields and very large values (e.g. $k=20$) lead to excessive smoothing.  Our conclusions are not limited to the specific choice of $k=5$. We find that all qualitative trends reported in this work remain unchanged within the range $3\leq k \leq7$.

To mitigate boundary-related biases in the estimation of local density, we explicitly account for the proximity of galaxies to the survey edges. For each galaxy, we compute the distance $r_b$ to the nearest survey boundary and retain only those galaxies satisfying $r_b > r_5$, ensuring that the fifth nearest-neighbour sphere is fully contained within the survey volume. We have verified that adopting a more conservative buffer ($r_b > 1.5\,r_5$) does not alter any of our qualitative results, but substantially reduces the available sample size and increases statistical uncertainties. We therefore adopt the $r_b > r_5$ criterion as an optimal balance between robustness against edge effects and statistical power.

The fifth-nearest neighbour density $\rho_5$ thus serves as a proxy for local overdensity, allowing us to disentangle the effects of small-scale interactions and immediate surroundings from the broader cosmic web environment. Combined with our Hessian-based classification of large-scale structure, this local density measure forms a crucial part of our framework for analyzing how environment shapes galaxy evolution.

While the smoothed CIC density field plays a central role in deriving the Hessian tensor and identifying large-scale cosmic web structures, it is not ideally suited for characterizing the immediate environment of individual galaxies. The CIC method relies on interpolating the galaxy or mass distribution onto a regular grid and applying a smoothing kernel over several megaparsecs ($8\,\mathrm{Mpc}$ in our case), typically to suppress shot noise and extract coherent large-scale features. As such, it captures the geometry of the cosmic web but washes out small-scale variations that are critical for understanding localized environmental effects.

In contrast, the fifth-nearest neighbour density, $\rho_5$, provides a direct, adaptive measure of local density on scales more relevant to galaxy-galaxy interactions, group environments, and pre-processing effects. Unlike CIC-based densities, which are sensitive to grid resolution and the choice of smoothing scale, $\rho_5$ dynamically adjusts to the local galaxy distribution, offering higher spatial resolution in dense regions and broader coverage in sparse ones.

Moreover, $\rho_5$ reflects the conditions experienced by galaxies on physical scales comparable to those over which processes like tidal interactions, ram-pressure stripping, and gas accretion operate. This makes it particularly well-suited for probing the connection between quenching and the immediate surroundings of galaxies. Thus, our use of $\rho_5$ allows for a more precise control over local environmental effects in our analysis, enabling a cleaner disentanglement of small-scale density-driven processes from the large-scale influence of the cosmic web topology.

\subsection{Stellar mass matching and density control}

\begin{figure*}[ht]
    \centering
    \includegraphics[width=0.65\textwidth]{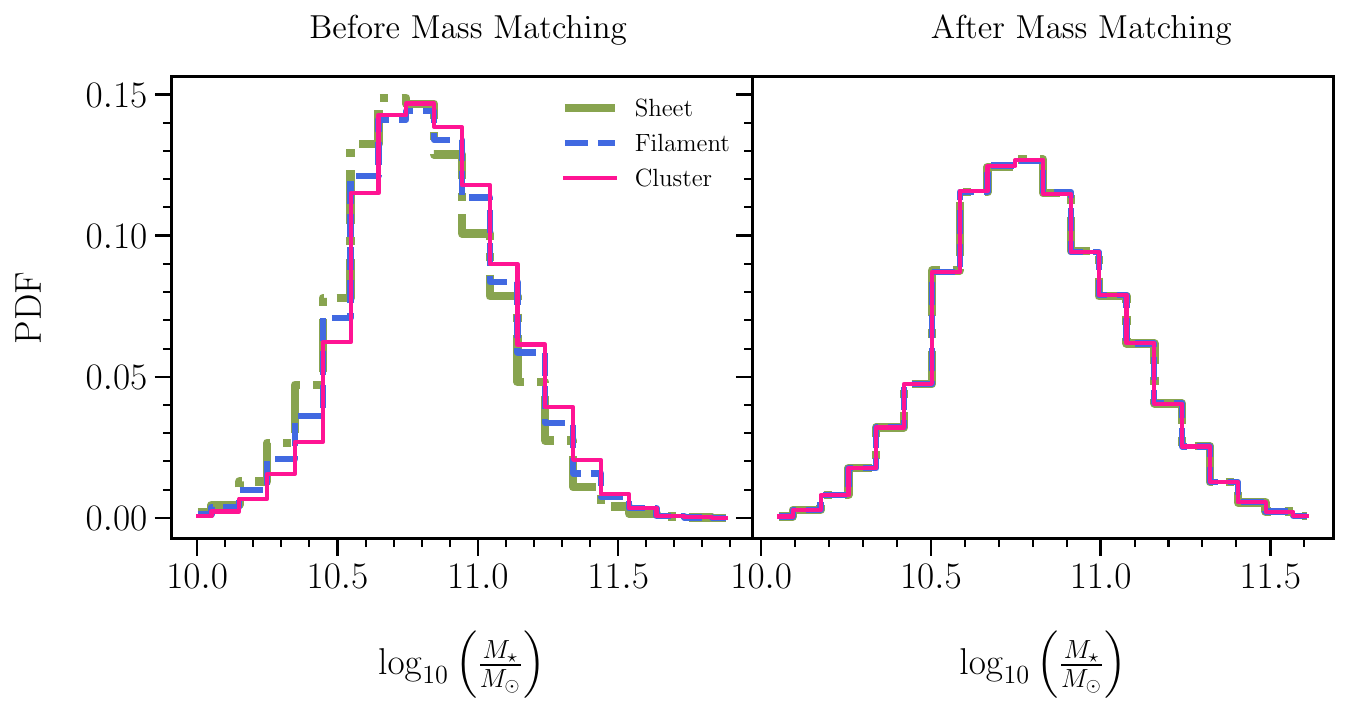}
    \caption{The left panel of this figure shows stellar mass distributions of galaxies in sheet, filament, and cluster environments. The distributions after stellar mass matching are shown in the right panel.}
    \label{fig:mass_matching}
\end{figure*}

To isolate the effect of environment from that of stellar mass, we perform stellar mass matching across sheets, filaments, and clusters. This ensures identical stellar mass distributions among galaxies in the three environments. As shown in \autoref{fig:mass_matching}, the unmatched distributions differ significantly, while the matched distributions are nearly identical. Post-matching, we retain 14,339 galaxies in each environment.

In contrast, we do not perform a strict local density matching. Instead, we restrict our analysis to galaxies within a common range of fifth nearest neighbour densities ($\rho_5$), defined by the overlapping region of the $\rho_5$ distributions across the three environments. This approach removes extreme density outliers and ensures that all comparisons are made within a shared density domain. However, it does not eliminate all residual differences in the internal $\rho_5$ distributions. As such, our analysis primarily isolates the effects of cosmic web topology at fixed stellar mass, while controlling local density only approximately. The resulting dataset is volume-limited, stellar mass-matched, and local density-controlled. This design enables robust and unbiased comparisons of galaxy properties across cosmic web environments, providing a clean framework to explore the differential roles of internal and external processes in galaxy quenching.


\section{Results}
\label{sec:results}

In this section, we present a comparative analysis of galaxy properties across distinct cosmic web environments such as sheets, filaments, and clusters focusing on quenched fraction, bulge fraction, and AGN fraction. To build a more complete picture of galaxy evolution, we further examine trends in specific star formation rate, $(u-r)$ colour, concentration index, and D4000 strength across the stellar mass–density plane. All results are based on a volume-limited, stellar mass-matched, and density-controlled galaxy sample drawn from SDSS DR18, ensuring a consistent and unbiased comparison across environments.

\subsection{Quenching, morphology, and AGN activity across cosmic web environments}
\begin{figure*}[ht]
    \centering
    \includegraphics[width=0.95\textwidth]{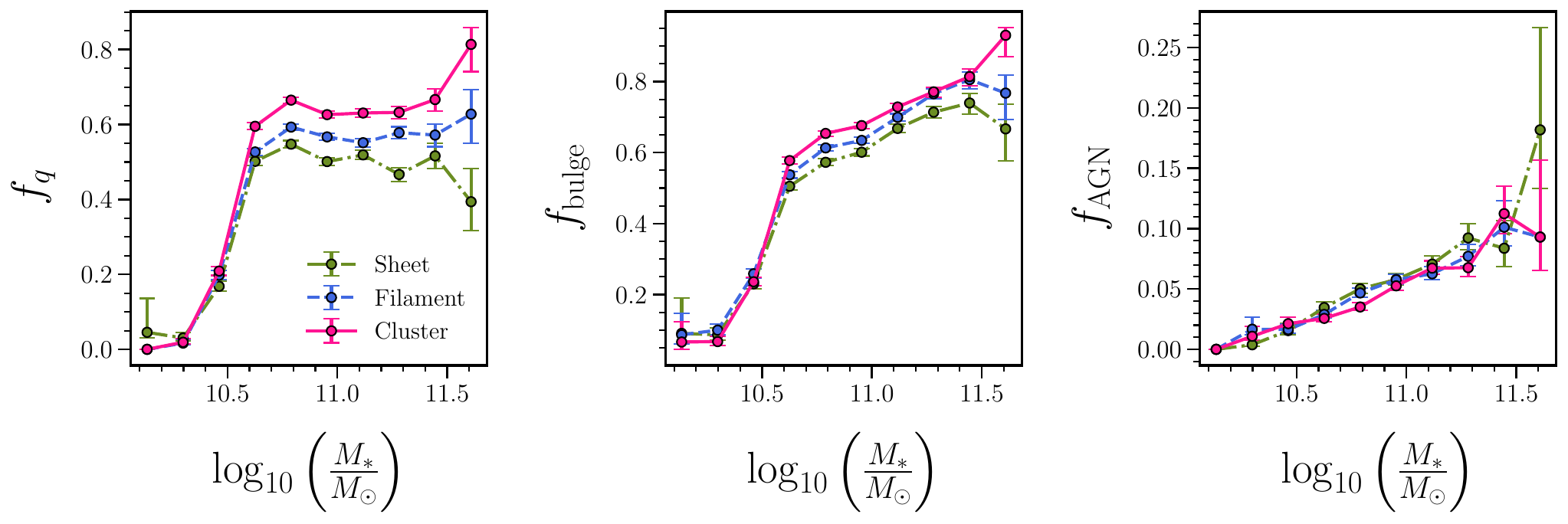}
    \caption{This shows the quenched fraction ($f_q$) (left panel), bulge fraction ($f_{\rm bulge}$) (middle panel), and AGN fraction ($f_{\rm AGN}$) (right panel) as functions of stellar mass for galaxies in sheet, filament, and cluster environments. Error bars show 1$\sigma$ uncertainties using beta distribution quantile technique. A flattening in the quenched fraction is seen beyond $\log_{10}(M_*/M_\odot) \sim 10.6$, followed by a bifurcation at higher mass where cluster galaxies continue quenching, but those in sheets exhibit a decline, potentially indicating gas re-accretion. Notably, while the quenched fraction flattens beyond $\log_{10}(M_\star/M_\odot) \sim 10.6$ in all environments, the bulge fraction continues to rise with stellar mass, highlighting a decoupling between star formation quenching and morphological transformation. At the high-mass end ($\log_{10}(M_\star/M_\odot) \gtrsim 11.5$), the bulge fraction bifurcates continuing to rise in clusters while declining in sheets revealing divergent evolutionary pathways in different cosmic web environments.}
\label{fig:fractions_plot}
\end{figure*}

\autoref{fig:fractions_plot} presents the quenched fraction ($f_q$), bulge fraction ($f_{\rm bulge}$), and AGN fraction ($f_{\rm AGN}$) as functions of stellar mass for galaxies residing in sheets, filaments, and clusters. The errorbars in \autoref{fig:fractions_plot} are estimated using the beta distribution quantile technique~\citep{cameron11}. At fixed stellar mass, galaxies in denser environments particularly clusters exhibit systematically higher quenched and bulge fractions compared to their counterparts in filaments and sheets. This hierarchy reflects the increasing influence of environmental quenching mechanisms such as ram-pressure stripping, tidal interactions, and gas strangulation within progressively denser regions of the cosmic web. 

A noticeable flattening in the quenched fraction occurs beyond $\log_{10}(M_\star/M_\odot) \sim 10.6$ (left panel of \autoref{fig:fractions_plot}), indicating a transition from environment-driven quenching at lower masses to mass-driven processes at higher masses. However, while the quenched fraction saturates, the bulge fraction (middle panel of \autoref{fig:fractions_plot}) continues to rise beyond this threshold. Although several observational studies indicate that transformation of galaxy morphology is closely linked to quenching in galaxies \citep{schawinski14, tacchella15, kawin17}, the trend observed in our analysis suggests that morphological transformation remains active even after star formation has largely ceased. This decoupling between quenching and structural evolution implies that processes such as mergers, internal disk instabilities, and bar-driven inflows may continue to reshape galaxies after the cessation of star formation. The rising bulge fraction at high masses, consistent across all environments, supports the idea that morphological quenching becomes increasingly important, even when other quenching mechanisms may have already acted. These observations reinforce that quenching and bulge growth, while correlated, are not causally synchronous. They evolve on different timescales and respond differently to internal and external conditions.

At the high-mass end ($\log_{10}(M_\star/M_\odot) \gtrsim 11.5$), we observe a striking bifurcation: while both quenched and bulge fractions continue to increase in clusters, they decline in sheets. We find that the differences observed in the quenched fraction and bulge fraction between sheet and cluster environments in the highest mass bin are statistically significant, with a confidence level of $\sim 99\%$ (\autoref{tab:ztest}). Moreover, this environmental divergence is not confined to the highest mass bin alone. Across the stellar mass range $10.5 \leq \log_{10}(M_\star/M_\odot) \leq 11.5$, the quenched fractions in sheet and cluster environments remain statistically distinct at a confidence level of $\gtrsim 99.99\%$, indicating a persistent and highly significant environmental imprint. Over the same mass range, the bulge fractions also differ significantly between sheets and clusters, with confidence levels ranging from $\sim95\%$ to $\sim99\%$. This parallel divergence suggests that massive galaxies in sheets follow a fundamentally different evolutionary trajectory. The lack of strong external quenching mechanisms such as high-speed encounters or ram-pressure stripping allows these galaxies to retain cold gas and disk-like morphologies, enabling sustained or rejuvenated star formation. The simultaneous suppression of quenching and bulge formation in sheets points to a population of massive, morphologically less transformed galaxies that remain actively evolving under favourable environmental conditions. Together, these results demonstrate that the bifurcation observed at the high-mass end represents the culmination of a broader, statistically robust divergence in both star-formation suppression and structural transformation across cosmic web environments. This scenario highlights the role of the cosmic web not merely as a static framework, but as an active agent influencing gas dynamics and star formation pathways.

AGN activity, as traced by $f_{\rm AGN}$ in the right panel of \autoref{fig:fractions_plot}, increases steadily with stellar mass across all environments. In contrast to the flattened or diverging trends seen in $f_q$ and $f_{\rm bulge}$, the AGN fraction continues rising even at the high-mass end. Intriguingly, at fixed stellar mass, AGN fractions are somewhat higher in sheets than in clusters particularly among massive galaxies. This suggests that while AGN fueling is governed primarily by internal processes such as central gas inflow and black hole accretion, it is also modulated by environmental conditions that determine gas availability. In dense cluster environments, quenching mechanisms may suppress or expel the gas reservoir needed to power AGN, whereas in the more benign conditions of sheets, galaxies may retain or reaccrete gas, facilitating prolonged or episodic AGN activity. The elevated AGN fraction in sheets coexisting with declining quenched and bulge fractions implies that AGN activity is not necessarily a terminal phase of galaxy evolution, but may coincide with or even stimulate ongoing star formation under the right conditions. 

Taken together, these trends reveal a nuanced interplay between stellar mass, morphology, AGN activity, and cosmic web environment. They underscore the importance of considering not only local density but also large-scale cosmic structure as a key regulator of gas availability, feedback processes, and galaxy evolution.

\begin{deluxetable}{lccc}
\tablecaption{Results of two-sample two-proportion $z$-tests assessing whether the observed differences in quenched, bulge-dominated, and AGN fractions between sheet and cluster environments in the highest mass bin are statistically significant.}\label{tab:ztest}
\tablewidth{0pt}
\tablehead{
\colhead{Environment Pair} &
\colhead{Property} &
\colhead{$z$-score} &
\colhead{$p$-value}
}
\startdata
Sheet-Cluster & Quenched fraction & $-3.76$ & $2.0\times10^{-4}$ \\
Sheet-Cluster & Bulge fraction    & $-2.94$ & $3.3\times10^{-3}$ \\
Sheet-Cluster & AGN fraction      & $1.14$  & $0.2564$ \\
\enddata
\end{deluxetable}

\begin{deluxetable}{lcccccc}
\tablecaption{Comparison of galaxy properties between sheet and cluster environments in the highest stellar-mass and lowest local-density bin (with $10 \times 10$ bins) of the mass-density plane. We report median values for sheet and cluster galaxies separately, along with the corresponding Mann-Whitney $U$ statistics and $p$-values.}
\label{tab:utest}
\tablewidth{0pt}
\tablehead{
\colhead{Property} &
\multicolumn{2}{c}{Median value} &
\multicolumn{2}{c}{$U$ statistic} &
\colhead{$p$-value} \\
\cline{2-3} \cline{4-5}
& \colhead{Sheet} & \colhead{Cluster}
& \colhead{Sheet} & \colhead{Cluster} &
}
\startdata
$\log_{10}({\rm sSFR/yr}^{-1})$ & $-10.80$ & $-22.47$ & $452$ & $156$ & $0.0041$ \\
$(u-r)$ colour                 & $2.56$   & $2.76$   & $186$ & $422$ & $0.0221$ \\
$r_{90}/r_{50}$                & $2.83$   & $3.08$   & $193$ & $415$ & $0.0313$ \\
D4000                          & $1.78$   & $1.91$   & $183$ & $425$ & $0.0189$ \\
\enddata
\end{deluxetable}

\subsection{Star formation in the mass-density plane}

We disentangle the influence of local environment from that of large-scale cosmic web geometry by analyzing galaxy properties directly in the two-dimensional stellar mass-local density plane for each cosmic web environment. Rather than enforcing an explicit local-density matching across sheets, filaments, and clusters which would severely reduce sample sizes and preferentially select galaxies from narrow overlap regions, we compare galaxies at fixed stellar mass and fixed local density through a common binning scheme. This approach preserves statistical power while allowing a controlled assessment of how galaxies occupying the same region of the mass-density plane behave differently depending on their position within the cosmic web. We adopt a fine binning strategy that balances number statistics against sensitivity to sharp environmental transitions, enabling us to capture distinct trends at the high-mass and low-density extremes that would otherwise be diluted by coarser binning.

\begin{figure*}
    \centering
    \includegraphics[width=0.65\textwidth]{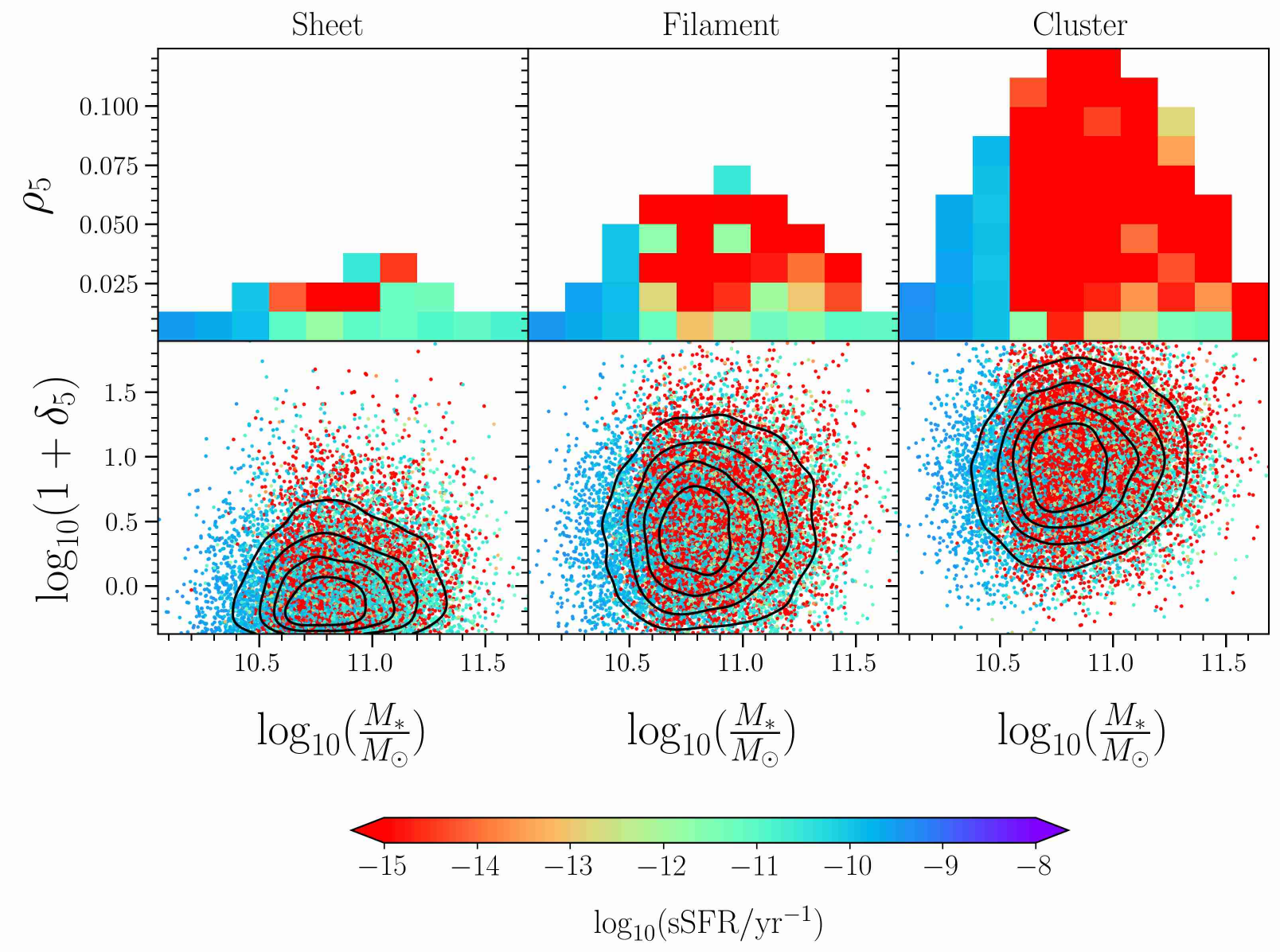}
    \caption{The top panel shows the median $\log_{10}(\mathrm{sSFR}/\mathrm{yr}^{-1})$ in the $\rho_5$-$\log_{10}(M_*/M_\odot)$ plane for sheet, filament, and cluster environments. The median $\log_{10}(\mathrm{sSFR}~\mathrm{yr}^{-1})$ are computed in $10 \times 10$ bins of $\log_{10}(M_*/M_\odot)$ and $\rho_5$. Only bins containing at least 10 galaxies are shown here. Notably, in the lowest-density, highest-mass bin, galaxies in sheets exhibit substantially higher median sSFR than those in clusters, emphasizing a clear environmental divergence in residual star formation. The bottom panel shows the individual galaxies plotted in the $\log_{10}(1+\delta_5)$-$\log_{10}(M_*/M_\odot)$ plane, colour-coded by sSFR, with iso-density contours overlaid.}
\label{fig:mass_density_ssfr}
\end{figure*}

To further explore how quenching is modulated by both stellar mass and local environment, we examine the distribution of specific star formation rates (sSFR) in the $\rho_5$-$\log_{10}(M_\star/M_\odot)$ and $\log_{10}(1 + \delta_5)$-$\log_{10}(M_\star/M_\odot)$ planes, where $\delta_5$ is the local density contrast defined as $\delta_5 = \rho_5/\bar{\rho} - 1$. \autoref{fig:mass_density_ssfr} provides a two-level visualization of sSFR behaviour across cosmic web environments. The upper panels show the median $\log_{10}({\rm sSFR/yr}^{-1})$ in $10 \times 10$ bins of $\rho_5$ and stellar mass, computed for cells containing at least 10 galaxies. The lower panels present corresponding scatter plots in the $\log_{10}(1+\delta_5)$-$\log_{10}(M_\star/M_\odot)$ plane, with each galaxy colour-coded by its sSFR and four iso-density contours overlaid to highlight structural trends.

Consistent with the quenched fraction trends in \autoref{fig:fractions_plot}, sSFR declines with both increasing stellar mass and local density across all environments. However, the strength and spatial extent of this suppression differ significantly by environment. Clusters exhibit widespread quenching across the entire mass-density plane, including intermediate masses and densities, indicating the joint influence of mass quenching and strong environmental effects such as ram-pressure stripping and strangulation. Filaments show more moderate suppression: although sSFR decreases with both variables, a substantial population of star-forming galaxies persists, particularly at intermediate densities. In sheets, sSFR remains elevated over the largest area of the parameter space, especially at low $\rho_5$ and $\delta_5$. Strikingly, even some massive galaxies ($\log_{10}(M_\star/M_\odot) \gtrsim 11$) in sheets retain relatively high sSFRs, suggesting that star formation is either not fully quenched or has been reactivated via gas re-accretion.

To directly assess whether the cosmic web environment exerts an influence on star formation beyond local density and stellar mass, we performed a targeted statistical comparison of galaxies residing in sheets and clusters within the highest stellar mass and lowest local density regime of the mass-density plane. Using the $10 \times 10$ binning adopted throughout this work, we identified galaxies occupying the same mass-density bin (highest mass and lowest density bins) but embedded in different large-scale environments, thereby isolating geometric effects associated with the cosmic web.

Given the strongly non-Gaussian nature of the sSFR distributions, we employed the non-parametric Mann-Whitney $U$ test to evaluate whether the two samples are drawn from the same parent distribution. This analysis reveals a statistically significant difference between sheet and cluster galaxies ($U_{\rm sheet}=452$, $U_{\rm cluster}=156$, $n_{\rm sheet}=32$, $n_{\rm cluster}=19$, $p=0.0041$, \autoref{tab:utest}). In particular, galaxies in sheets exhibit systematically higher star formation activity, with a median $\log_{10}(\mathrm{sSFR}/\mathrm{yr}^{-1})=-10.80$, compared to a dramatically suppressed median value of $\log_{10}(\mathrm{sSFR}/\mathrm{yr}^{-1})=-22.47$ for galaxies in clusters. This result provides direct statistical evidence that, at fixed stellar mass and local density, the cosmic web environment significantly modulates star formation efficiency.

\begin{figure*}
\centering
\includegraphics[width=0.65\textwidth]{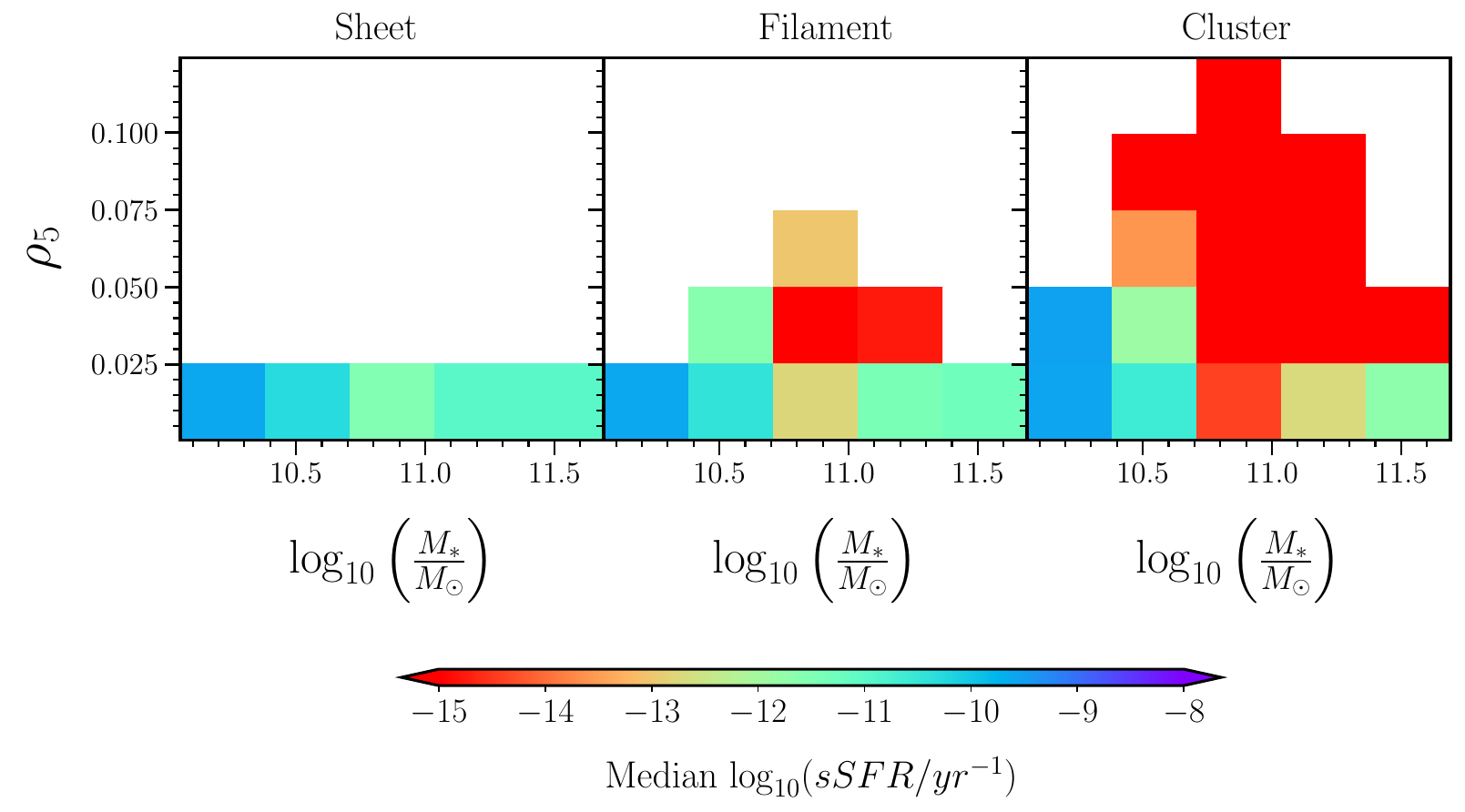}
\caption{This shows the distribution of $\log_{10}(\mathrm{sSFR/yr}^{-1})$ in the stellar mass-local density plane for sheet (left), filament (middle), and cluster (right) environments using a $5 \times 5$ binning scheme. Each cell represents the median sSFR of galaxies within that mass-density bin. Only bins containing at least 40 galaxies are retained to ensure statistical reliability. The enhanced visual contrast in the coarser binning highlights the systematic suppression of star formation in clusters relative to sheets at fixed stellar mass and local density. The environmental trends remain consistent with those obtained using the fiducial $10 \times 10$ binning (\autoref{fig:mass_density_ssfr}), demonstrating that our conclusions are robust to binning resolution.}
\label{fig:mass_density_ssfr_5x5}
\end{figure*}

To test the robustness of this conclusion against binning choices, we repeated the analysis using a coarser $5 \times 5$ binning scheme in the mass-density plane (\autoref{fig:mass_density_ssfr_5x5}). Although this approach increases the sample size within the highest mass and lowest density bin ($n_{\rm sheet}=244$, $n_{\rm cluster}=209$) and partially smooths environmental contrasts, the difference in sSFR between sheets and clusters still remains statistically significant ($U_{\rm sheet}=28765.5$, $U_{\rm cluster}=22230.5$, $p=0.0187$). The corresponding median sSFR values, $\log_{10}(\mathrm{sSFR}/\mathrm{yr}^{-1})=-10.99$ for sheets and $-11.69$ for clusters, continue to demonstrate enhanced star formation in sheet environments. Thus the same environmental divergence is clearly visible in the $5 \times 5$ representation of the mass-density plane (\autoref{fig:mass_density_ssfr_5x5}), and remains statistically significant under a Mann-Whitney $U$ test ($p=0.0187$).

Taken together, these results quantitatively reinforce our visual and statistical analyses of the mass-density plane, demonstrating that the elevated sSFR of massive galaxies in sheets is not an artifact of binning or sampling, but a genuine signature of cosmic web driven regulation of star formation.

Several key insights emerge from the patterns observed in \autoref{fig:mass_density_ssfr}. First, galaxies with $\log_{10}(M_\star/M_\odot) \lesssim 10.5$ are predominantly star-forming in all environments, with the most active systems occupying low-density regions. Second, at fixed stellar mass, galaxies in sheets consistently exhibit higher median sSFR than those in filaments or clusters, particularly at low densities. This points to a reduced efficiency of quenching in sheets and highlights the role of cosmic web geometry in governing the thermodynamic state of gas. Third, the persistence of star-forming activity among massive galaxies in sheets supports the possibility of rejuvenated star formation enabled by lower ambient pressure and coherent, filament-fed inflows. Conversely, the near-complete suppression of sSFR in clusters underscores the importance of external mechanisms that disrupt gas inflow and heating.

These findings, viewed alongside the bifurcation in quenched and bulge fractions at high stellar mass (\autoref{fig:fractions_plot}), reinforce a picture in which galaxy evolution is modulated not only by mass but also by environment. While mass sets a global threshold for quenching, the geometry of the cosmic web and the physical state of the gas determine whether that threshold is crossed, and whether quenching is permanent or reversible. Sheet environments, in particular, appear capable of sustaining star formation in massive galaxies that would otherwise be quenched in denser regions. Together, these results suggest that the cosmic web is not merely a passive structure but an active regulator of star formation and morphological transformation.

\subsection{Colour, morphology and stellar age in the mass-density plane}

\begin{figure*}[ht]
    \centering
    \includegraphics[width=0.65\textwidth]{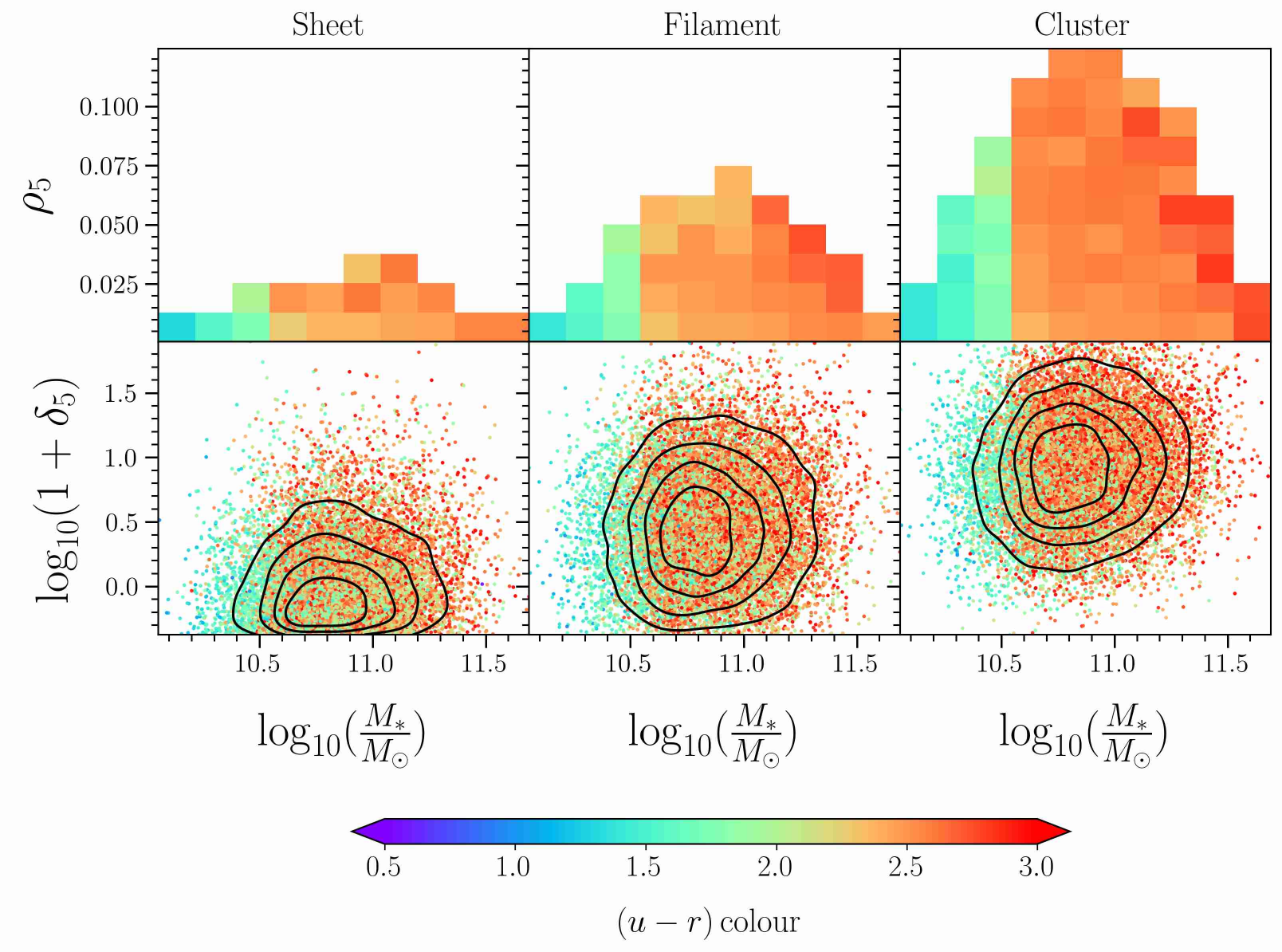}
    \caption{Same as \autoref{fig:mass_density_ssfr} but for $(u-r)$ colour. In the lowest-density, highest-mass regime, galaxies in sheets remain significantly less redder than their cluster counterparts, reflecting sustained or rejuvenated star formation in low-density environments.}
\label{fig:mass_density_color}
\end{figure*}

\begin{figure*}[ht]
    \centering
    \includegraphics[width=0.65\textwidth]{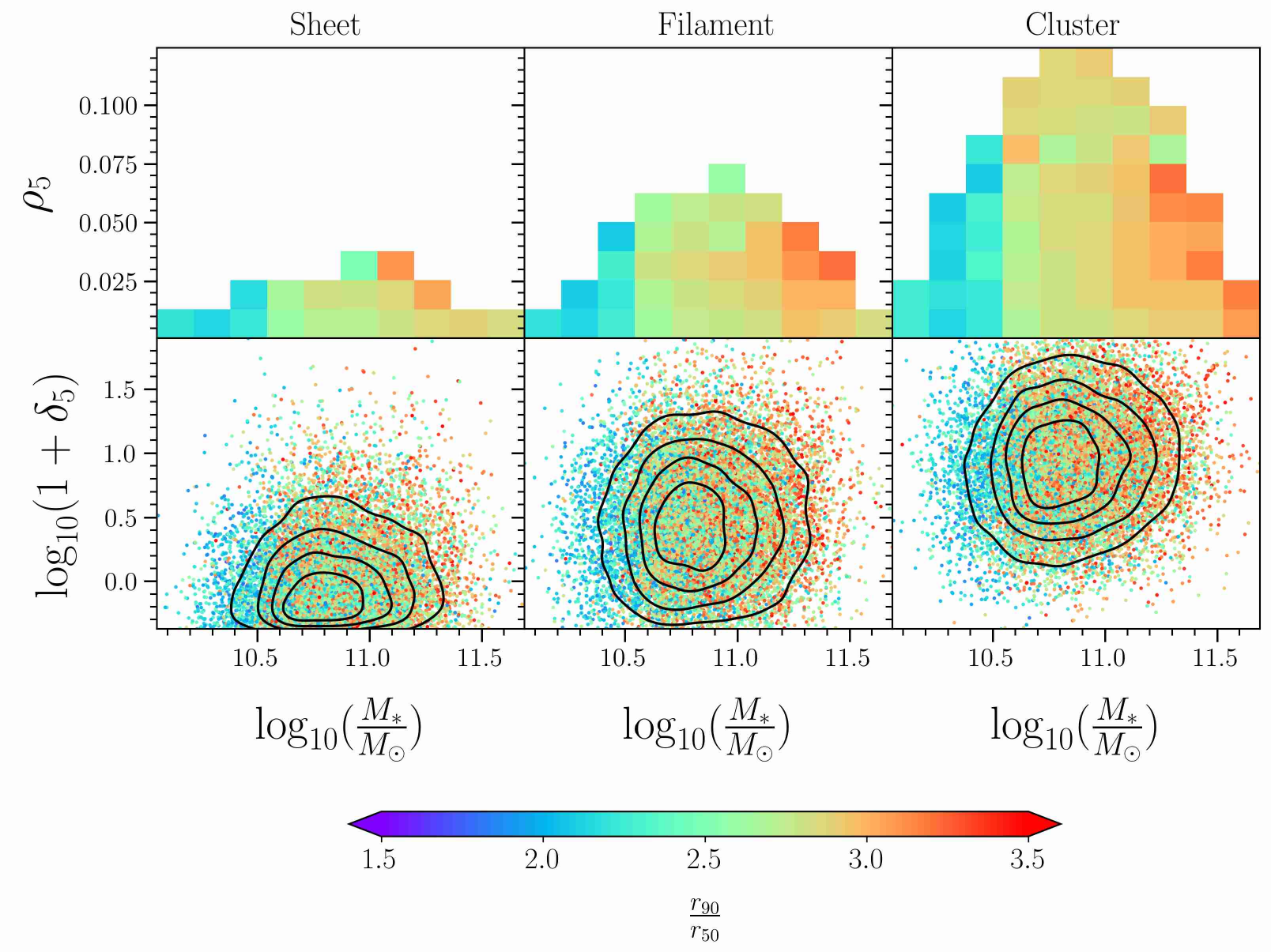}
    \caption{Same as \autoref{fig:mass_density_ssfr} but for concentration index ($r_{90}/r_{50}$). A striking difference emerges in the highest-mass, lowest-density bin, where galaxies in sheets remain less concentrated than those in clusters, signaling a lag in morphological transformation.}
\label{fig:mass_density_ci}
\end{figure*}

\begin{figure*}[ht]
    \centering
    \includegraphics[width=0.65\textwidth]{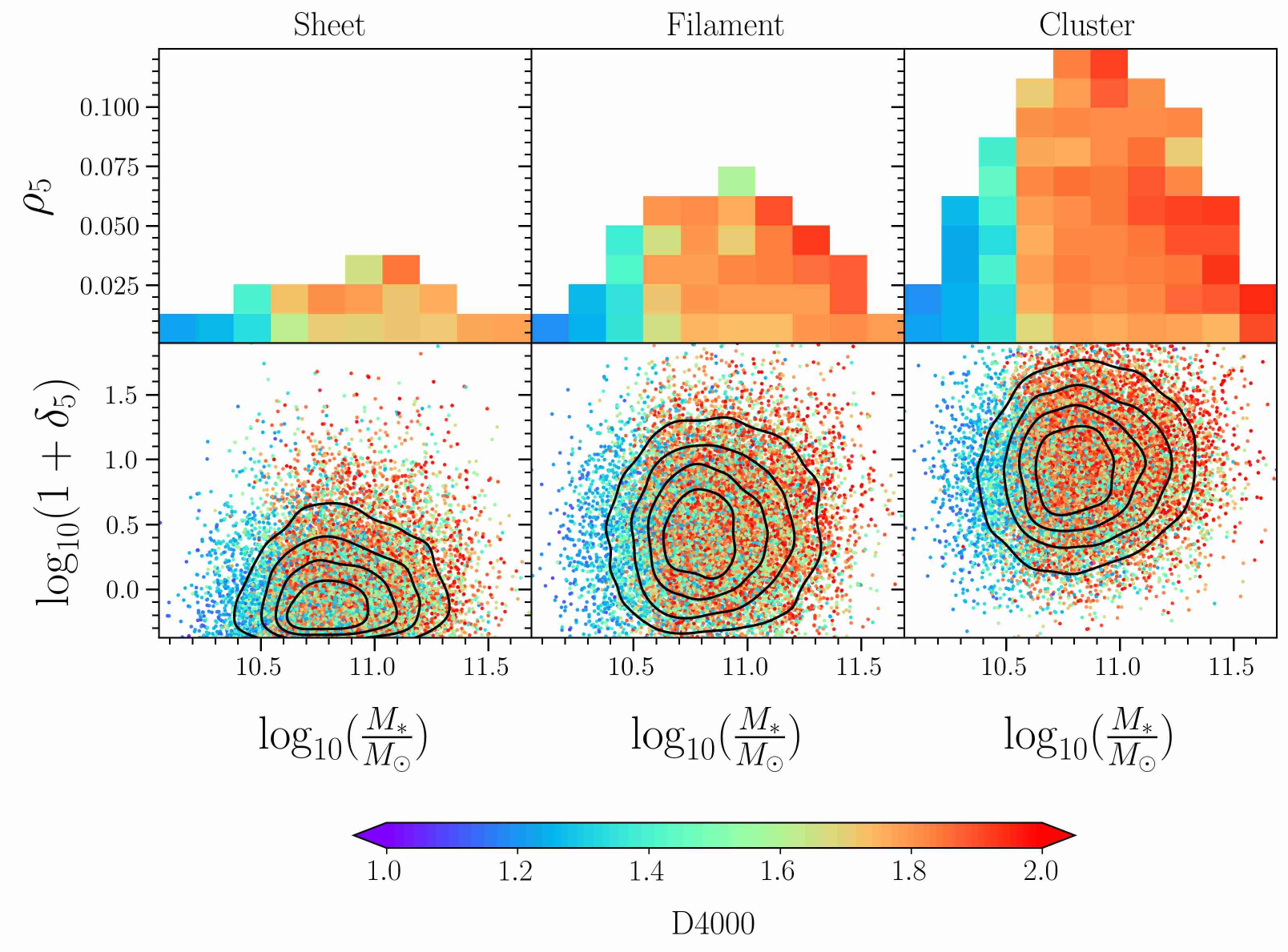}
    \caption{Same as \autoref{fig:mass_density_ssfr} but for D4000. In the lowest-density, highest-mass bin, galaxies in sheets show lower D4000 values compared to those in clusters, pointing to younger stellar populations and delayed quenching.}
\label{fig:mass_density_d4}
\end{figure*}

To further characterize the quenching and evolutionary state of galaxies across environments, we analyze three key observables $(u-r)$ colour, concentration index ($r_{90}/r_{50}$), and D4000 break strength in the $\rho_5$-$\log_{10}(M_\star/M_\odot)$ and $\log_{10}(1+\delta_5)$-$\log_{10}(M_\star/M_\odot)$ planes (\autoref{fig:mass_density_color}, \autoref{fig:mass_density_ci}, and \autoref{fig:mass_density_d4}). Together, these diagnostics provide a multifaceted view of colour, morphology, and stellar population age across sheets, filaments, and clusters.

In addition to sSFR, \autoref{tab:utest} demonstrates that quenched galaxies in sheets and clusters differ significantly in their stellar populations and structural properties, even when compared at fixed stellar mass and local density (highest mass and lowest density bin). Cluster galaxies are systematically redder than their sheet counterparts, with higher median $(u-r)$ colours, indicating more evolved stellar populations and a longer time since the cessation of star formation. Consistent with this picture, cluster galaxies also exhibit significantly higher D4000 values, reflecting older luminosity-weighted stellar ages and more permanent quenching histories. Structural differences parallel these trends: the concentration index is markedly higher in clusters, pointing to enhanced bulge dominance and more advanced morphological transformation. In contrast, galaxies in sheets retain lower concentrations, bluer colours, and younger stellar populations, suggesting delayed or incomplete quenching despite their high stellar masses. The statistical significance of these differences, quantified by Mann-Whitney $U$ tests, confirms that the divergence between sheet and cluster galaxies at the high-mass, low-density end extends beyond star formation activity to encompass stellar age and morphology, underscoring the role of large-scale cosmic web environment in shaping galaxy evolution.

\autoref{fig:mass_density_color} reveals that galaxy colours shift progressively redward with increasing stellar mass and local density, in agreement with the sSFR trends. Using the threshold $(u-r) = 2.22$ from \citet{strateva01} to separate blue and red populations, we find that clusters harbour a dominant red sequence even at intermediate masses and densities, while sheets preserve a substantial blue population over a wide parameter space. Filaments again exhibit intermediate behaviour, forming a bridge between the two extremes. The lower panels further emphasize this environmental segregation, with red galaxies concentrated in the dense peaks of clusters and bluer systems populating the low-density outskirts of sheets and filaments. These trends echo the picture of progressive environmental quenching driven by local conditions and cosmic web topology.

Morphological evolution is traced using the concentration index in \autoref{fig:mass_density_ci}, a proxy for bulge dominance. Galaxies in clusters consistently exhibit higher concentrations often exceeding the disk-bulge transition threshold of $r_{90}/r_{50} = 2.6$ \citep{strateva01} particularly at $\log_{10}(M_\star/M_\odot) \gtrsim 10.5$ and in denser environments. This reflects enhanced bulge growth via environmental processes such as mergers and tidal interactions. In contrast, galaxies in sheets largely retain disk-like morphologies, even at high masses and moderate densities. Filaments once again show a mixed population, with intermediate concentrations suggesting partial morphological transformation. These structural signatures align with the variations in sSFR and colour, reinforcing the idea that morphological evolution is closely linked to environmental quenching.

Stellar population ages, probed via the D4000 break strength (\autoref{fig:mass_density_d4}), show a similar environmental stratification. Median D4000 increases with both stellar mass and local density, reflecting accelerated aging in more massive and denser systems. Cluster galaxies display uniformly high D4000 values, indicating early and efficient quenching. Filaments show a more gradual rise in D4000, while sheets retain lower values across a wide mass and density range. Notably, even some massive galaxies in sheets exhibit intermediate D4000 values, suggesting recent or ongoing star formation, possibly aided by late-time gas re-accretion. These results offer compelling evidence that galaxy aging, like quenching and morphological change, is modulated by both local density and the broader cosmic web environment.

Altogether, the distributions of $(u-r)$, concentration index, and D4000 in the mass-density plane provide consistent and complementary support for our earlier conclusions, highlighting the nuanced and environment-dependent pathways of galaxy evolution.

\begin{figure*}
    \centering
    \includegraphics[width=0.65\textwidth]{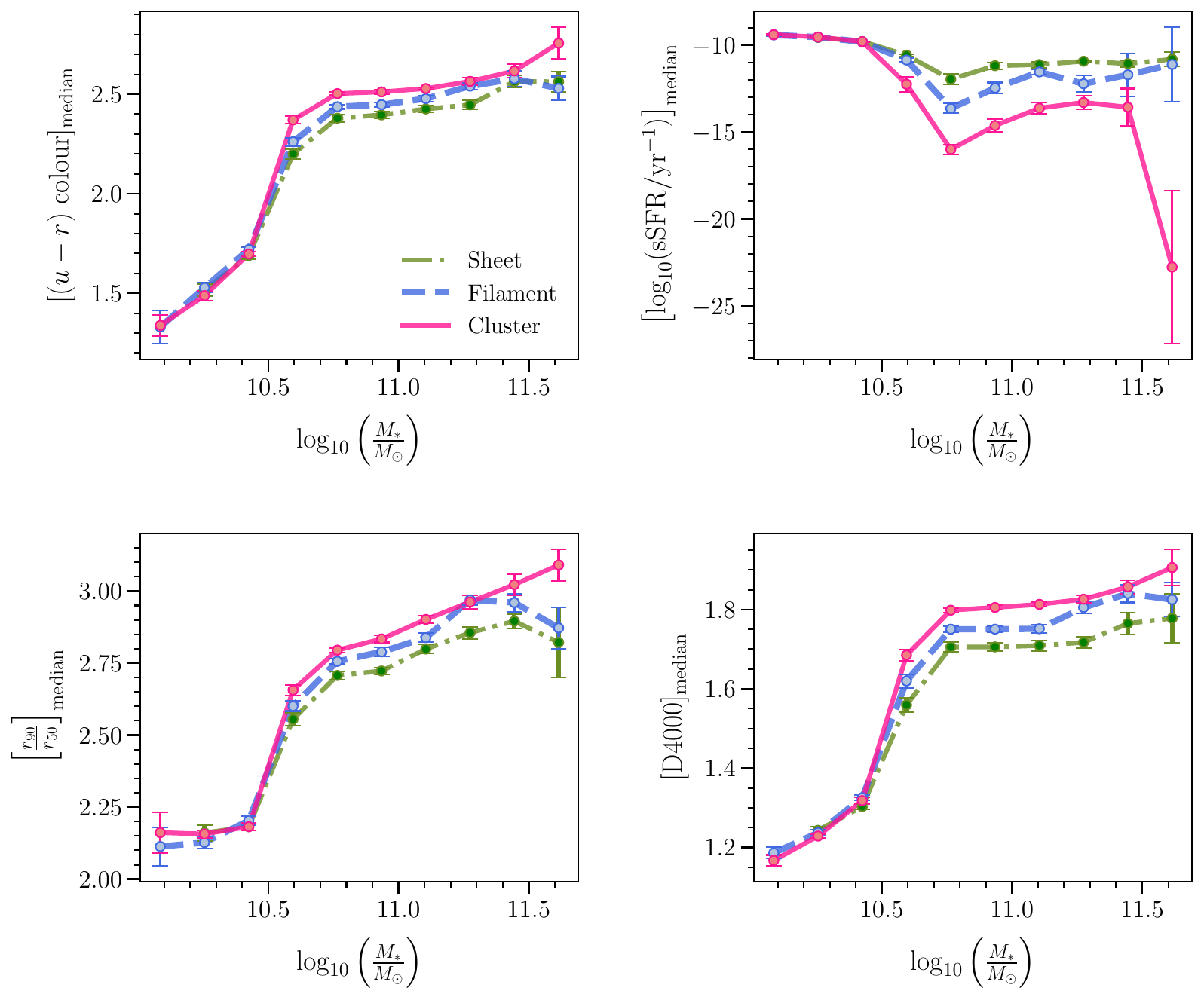}
\caption{This shows median $(u-r)$ colour (top left), $\log_{10}(\mathrm{sSFR}/\mathrm{yr}^{-1})$ (top right), concentration index $r_{90}/r_{50}$ (bottom left), and D4000 break strength (bottom right) as functions of stellar mass for galaxies in sheets, filaments, and clusters. The curves represent the medians within stellar mass bins, with $1\sigma$ scatter from 50 bootstrap realizations. Across all properties, galaxies in clusters appear redder, more quenched, structurally more concentrated, and older than their counterparts in sheets, with filaments displaying intermediate behaviour. Notably, the environmental distinctions sharpen at the high-mass end, pointing to an amplified role of the cosmic web in regulating galaxy evolution beyond the quenching threshold.}
\label{fig:medians_props_for_masked}
\end{figure*}

To further probe environmental influences, \autoref{fig:medians_props_for_masked} presents how key galaxy properties such as $(u-r)$ colour, $\log_{10}(\mathrm{sSFR}/\mathrm{yr}^{-1})$, concentration index ($r_{90}/r_{50}$), and D4000 systematically vary with stellar mass across different cosmic web environments, for the volume-limited, stellar mass-matched sample. The top panels show the median $(u-r)$ colour and $\log_{10}(\mathrm{sSFR})$ respectively, while the bottom panels display the concentration index ($r_{90}/r_{50}$) and the D4000 break strength. Across these panels, galaxies in clusters exhibit redder colours, more suppressed sSFR, higher concentration indices, and larger D4000 values compared to their counterparts in sheets and filaments, indicating more evolved stellar populations and greater bulge dominance. Sheets, by contrast, host galaxies that remain significantly bluer, less quenched, and morphologically disk-like, even at high stellar masses, reinforcing the idea of inefficient or reversible quenching in low-density environments. Filaments show intermediate trends, consistent with a transitional role in the quenching and morphological transformation processes. The divergence among environments becomes especially pronounced at the highest stellar masses, suggesting that environmental modulation becomes increasingly important beyond the quenching threshold. These systematic differences across environments lend additional support to our earlier findings, highlighting the profound influence of cosmic web topology on the co-evolution of star formation, colour, morphology, and stellar population age.

\begin{figure*}
    \centering
    \includegraphics[width=0.65\textwidth]{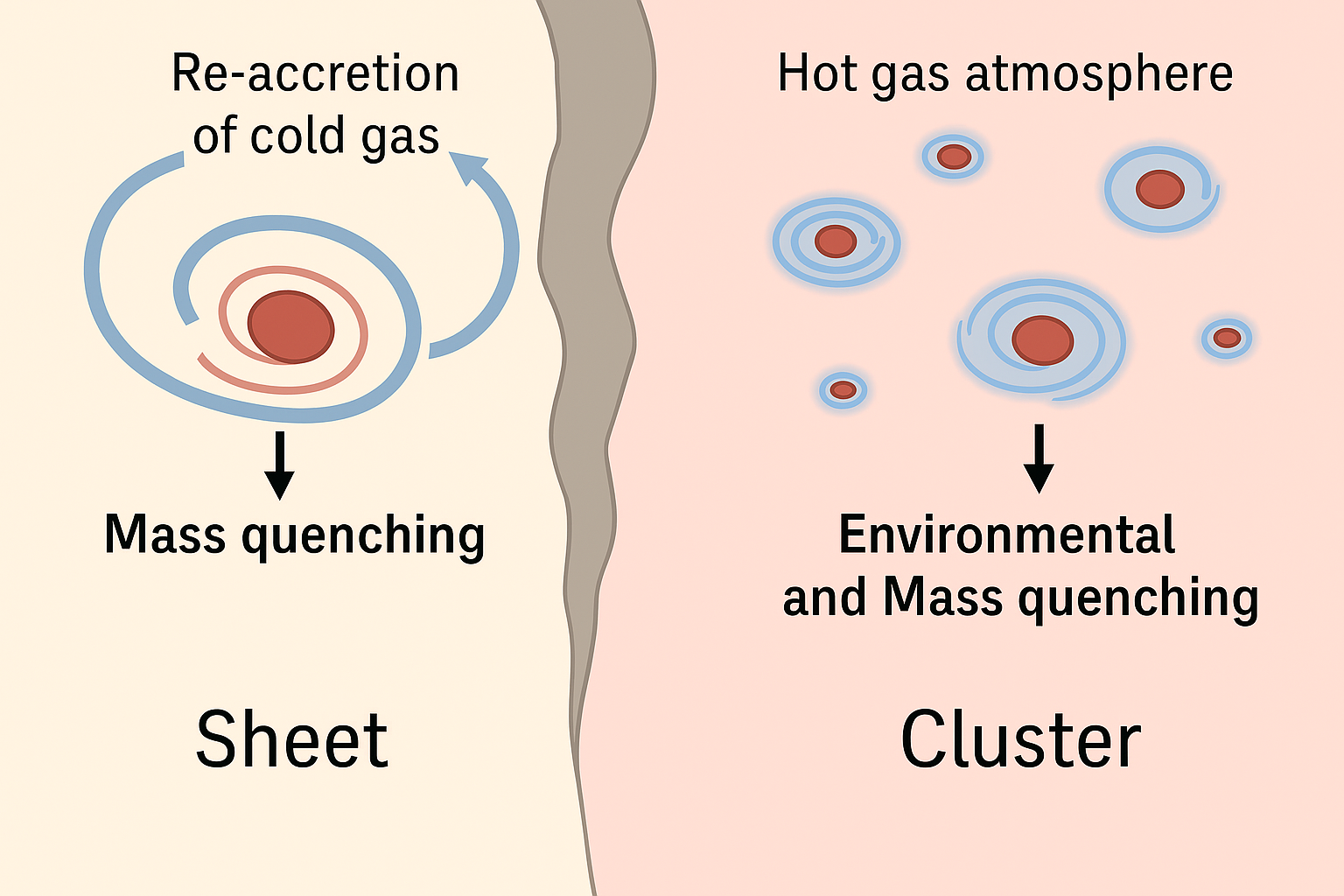}
    \caption{This schematic illustration shows the quenching pathways across cosmic web environments. Clusters drive strong, irreversible quenching via environmental and internal processes. In sheets, mass quenching dominates, and re-accretion may revive star formation. Filaments lie in between, with mixed influence.}
    \label{fig:interpretation}
\end{figure*}

\autoref{fig:interpretation} provides a schematic view of the quenching pathways across the cosmic web. In clusters, both environmental and mass quenching work in tandem, leading to rapid, irreversible suppression of star formation. In sheets, mass quenching alone dominates, and low ambient pressure may permit late-time gas inflow, particularly for massive galaxies. Filaments represent a transitional regime.

These findings collectively support a model in which galaxy quenching is not only mass-dependent but also tightly regulated by large-scale structure. The cosmic web topology influences the thermal state, gas supply, and feedback retention of galaxies, ultimately shaping their star formation histories.

While the bifurcation in quenched and bulge fractions at the high-mass end ($\log_{10}(M_\star/M_\odot) \gtrsim 11.5$) (\autoref{fig:fractions_plot}) stands out as one of the most intriguing findings of this study, it is important to consider the sample sizes involved: among galaxies in this mass bin, 13 out of 33 are quenched in sheets, compared to 27 out of 43 in filaments and 35 out of 43 in clusters. Despite the modest statistics, the observed divergence, where quenched and bulge fractions continue to rise in clusters but decline in sheets, is consistently supported by other independent diagnostics such as the specific star formation rate, colour, concentration index and D4000. It may be noted that the galaxy counts quoted here differ from those reported for the highest stellar-mass and lowest density bin in the mass-density plane analysis that employs a $10\times10$ binning scheme and retains only cells containing at least 10 galaxies. Consequently, the highest-mass, lowest-density bin there represents a subset of the full high-mass sample rather than the entire population in that mass range. The coherence of the observed trends across multiple observables lends qualitative strength to the interpretation that massive galaxies in sheets may follow a distinct evolutionary trajectory, potentially characterized by weaker environmental quenching and enhanced capacity for gas retention or reaccretion. However, due to the smaller number of galaxies in the highest mass bin, this result should be interpreted with caution and warrants confirmation from larger datasets in future surveys. Nonetheless, the observed trend opens a valuable window into the potential role of cosmic web topology in regulating galaxy evolution at the highest masses.\\


\section{Conclusions}
\label{sec:conclusion}
In this study, we have investigated the role of the cosmic web in regulating galaxy quenching using a volume-limited sample of galaxies from SDSS DR18. Galaxies were classified into sheets, filaments, and clusters via a Hessian-based method that traces the curvature of the gravitational potential. To isolate the influence of environment, we constructed stellar mass-matched samples and restricted our analysis to a common range of local density, as measured by the fifth nearest neighbour ($\rho_5$). This framework enabled a controlled comparison of quenched fractions, morphological indicators, AGN activity, and specific star formation rates (sSFR) across distinct large-scale environments.

We find a clear environmental dependence in the quenched fraction: galaxies in clusters exhibit significantly higher quenching rates than their counterparts in filaments or sheets at fixed stellar mass. The quenched fraction increases with stellar mass up to $\log_{10}(M_\star/M_\odot) \sim 10.6$, beyond which it flattens, suggesting a transition from environment-driven quenching at low masses to mass-driven mechanisms at higher masses. In contrast, the bulge fraction continues to rise beyond this threshold, indicating that morphological transformation persists even after star formation has been largely suppressed. This divergence implies a decoupling of quenching and structural evolution, with processes such as mergers, bar instabilities, and internal disk dynamics driving bulge growth independently of star formation cessation.

At the high-mass end ($\log_{10}(M_\star/M_\odot) \gtrsim 11.5$), both quenched and bulge fractions exhibit a striking bifurcation: they continue to rise in clusters but decline in sheets. This coherent trend suggests a distinct evolutionary pathway in low-density environments. In sheets where external quenching mechanisms like ram-pressure stripping and galaxy harassment are weak, massive galaxies may retain or reacquire cold gas, sustain disk-like morphologies, and undergo rejuvenated star formation. This points to a scenario where galaxies in sheets are not merely delayed in their evolution but follow fundamentally different trajectories shaped by their cosmic web context.

Our analysis also shows that AGN fraction increases steadily with stellar mass in all environments, but reaches slightly higher values in sheets than in clusters at the high-mass end. This indicates that while AGN fueling is governed by internal factors such as gas inflow and black hole accretion, it is also modulated by environmental conditions that affect gas retention. In cluster environments, pervasive heating and gas removal may limit the available fuel for AGN activity. Conversely, the milder conditions in sheets enable the buildup and retention of cold gas, supporting sustained AGN fueling. The elevated AGN activity in sheets particularly in massive galaxies with declining quenched and bulge fractions suggests that AGN may not always mark the end of star formation but could instead coexist with, or even facilitate, intermittent star-forming episodes. These results highlight the importance of situating AGN within the broader ecosystem of environmental and internal processes that shape galaxy evolution.

An analysis of sSFR in the stellar mass-density plane further reinforces these conclusions. While sSFR declines with increasing mass and density in all environments, the suppression is most severe in clusters and weakest in sheets. Remarkably, even at high stellar masses ($\log_{10}(M_\star/M_\odot) \gtrsim 11$), galaxies in sheets maintain elevated sSFR in low-density regions, suggesting inefficient quenching or the possibility of late-time gas accretion. This persistence of star formation in sheets underscores the role of ambient pressure and gas inflow geometry in sustaining galaxy growth. Filaments exhibit intermediate behaviour, acting as a transitional environment where both quenching and accretion processes compete. These trends, when viewed in conjunction with the bifurcation in quenched and bulge fractions, provide strong evidence that the cosmic web actively shapes star formation and morphological transformation through its imprint on the thermal and dynamical state of circumgalactic gas.
This picture is further supported by the distributions of $(u-r)$ colour, concentration index, and D4000 for the full galaxy population, which show systematic shifts toward redder colours, higher central concentrations, and older stellar populations in denser environments. Notably, at the high-mass end ($\log_{10}(M_\star/M_\odot) \gtrsim 11.5$), these trends bifurcate: galaxies in clusters appear more evolved, while those in sheets remain bluer, younger, and morphologically less concentrated strengthening the case for divergent evolutionary pathways shaped by the cosmic web.

Taken together, our results demonstrate that galaxy quenching is a multi-faceted process governed by stellar mass, local density and large-scale environment. While clusters quench galaxies through the combined action of internal and external mechanisms, sheets offer a more permissive environment where massive galaxies can resist or even reverse quenching. This highlights the role of the cosmic web not merely as a structural scaffold but as a dynamic agent that regulates gas supply, feedback effectiveness, and structural evolution. Sheet environments, in particular, emerge as key laboratories for understanding how galaxies can remain actively evolving despite surpassing the canonical quenching thresholds.

Our results reinforce and expand upon recent findings from both simulations and observations that emphasize the pivotal role of the cosmic web in modulating galaxy quenching and morphology. Consistent with results from hydrodynamical simulations such as IllustrisTNG \citep{donnari21, hasan23, malavasi22} and Horizon-AGN \citep{song21}, we find that quenching is strongly environment-dependent, with clusters exhibiting widespread suppression of star formation across all masses, while filaments and sheets present more nuanced behaviour. In agreement with \citet{pallero19} and \citet{donnari21}, our results support a dual-mode picture in which environmental quenching dominates at low stellar mass, whereas mass quenching, likely linked to AGN feedback, becomes dominant above $\log_{10}(M_\star/M_\odot) \sim 10.5$. However, our analysis reveals an additional layer of complexity: at the high-mass end, galaxies in sheet environments display a bifurcation in quenched and bulge fractions not captured in earlier studies. This trend suggests that under certain geometric and thermodynamic conditions, massive galaxies in low-density environments can sustain or even rejuvenate star formation potentially due to residual or reaccreted cold gas. Furthermore, while studies such as \citet{okane24} argue that the environmental effects of filaments are largely attributable to local density, our findings suggest that the cosmic web topology itself particularly in sheet-like regions imposes additional regulatory effects not easily parameterized by local density alone. The elevated AGN activity we observe in massive sheet galaxies echoes predictions by \citet{kraljic20}, where favourable gas inflow conditions may facilitate black hole fueling. Our results also offer observational support for the cosmic web detachment model \citep{aragon19, maret20}, which posits that galaxies embedded in the cosmic web gradually decouple from their surrounding filaments as their host halos evolve, leading to a cessation of coherent cold gas inflow and eventual quenching. In this framework, galaxy quenching is intimately tied to the dynamics of their large-scale environment. Our finding that massive galaxies in sheet environments exhibit declining quenched and bulge fractions, even at $\log_{10}(M_\star/M_\odot) \gtrsim 11.5$, suggests that these galaxies remain attached to filamentary structures and can continue to re-accrete cold gas. In contrast, galaxies in clusters often located at cosmic nodes have likely undergone web detachment, leading to both efficient quenching and morphological transformation. This interpretation is further strengthened by a non-parametric statistical comparison at fixed stellar mass and local density, which shows that massive galaxies in sheets exhibit systematically higher sSFR, bluer $(u-r)$ colours, lower concentration indices, and smaller D4000 values than their cluster counterparts, confirming that cosmic web geometry modulates star formation, stellar populations, and structural evolution beyond the influence of local density alone. Further, \citet{bluck22} confirm that bulge mass is the single strongest predictor of quenching across a wide mass range supporting our scenario of ongoing morphological transformation decoupled from star-formation cessation. Altogether, our results build a more complete and observationally anchored picture of galaxy evolution within the cosmic web, highlighting the importance of environment-driven diversity not only in quenching mechanisms but also in the timing, morphology, and feedback pathways that shape galaxies across cosmic time.

Future work incorporating spatially resolved cold gas observations and high-resolution cosmological simulations will be crucial for disentangling the physical processes that drive these trends. Probing the thermodynamics, angular momentum, and inflow histories of galaxies in different cosmic web environments will shed further light on how large-scale structure influences galaxy evolution. Our findings affirm that the topology of the cosmic web is not just a passive backdrop, but a vital regulator of how galaxies live, quench, and transform across cosmic time.

\section{Acknowledgments}

We thank an anonymous reviewer and the statistics editor for their valuable comments and suggestions that helped us to significantly improve the draft. AN acknowledges the financial support from the Department of Science and Technology (DST), Government of India through an INSPIRE fellowship. BP would like to acknowledge IUCAA, Pune, for providing support through the associateship programme.

Funding for the Sloan Digital Sky Survey V has been provided by the Alfred P. Sloan Foundation, the Heising-Simons Foundation, the National Science Foundation, and the Participating Institutions. SDSS acknowledges support and resources from the Center for High-Performance Computing at the University of Utah. SDSS telescopes are located at Apache Point Observatory, funded by the Astrophysical Research Consortium and operated by New Mexico State University, and at Las Campanas Observatory, operated by the Carnegie Institution for Science. The SDSS web site is \url{www.sdss.org}.

SDSS is managed by the Astrophysical Research Consortium for the Participating Institutions of the SDSS Collaboration, including the Carnegie Institution for Science, Chilean National Time Allocation Committee (CNTAC) ratified researchers, Caltech, the Gotham Participation Group, Harvard University, Heidelberg University, The Flatiron Institute, The Johns Hopkins University, L'Ecole polytechnique f\'{e}d\'{e}rale de Lausanne (EPFL), Leibniz-Institut f\"{u}r Astrophysik Potsdam (AIP), Max-Planck-Institut f\"{u}r Astronomie (MPIA Heidelberg), Max-Planck-Institut f\"{u}r Extraterrestrische Physik (MPE), Nanjing University, National Astronomical Observatories of China (NAOC), New Mexico State University, The Ohio State University, Pennsylvania State University, Smithsonian Astrophysical Observatory, Space Telescope Science Institute (STScI), the Stellar Astrophysics Participation Group, Universidad Nacional Aut\'{o}noma de M\'{e}xico, University of Arizona, University of Colorado Boulder, University of Illinois at Urbana-Champaign, University of Toronto, University of Utah, University of Virginia, Yale University, and Yunnan University.

\section{Data availability}
The Sloan Digital Sky Survey (SDSS) data used in this study are publicly available through the SDSS CasJobs interface at \url{https://skyserver.sdss.org/casjobs/}. The derived data products generated in this work are available from the authors upon reasonable request.

\bibliography{quenching.bib}{}
\bibliographystyle{aasjournal}

\end{document}